\documentclass[letterpaper]{article} 
\usepackage{aaai24}  
\usepackage{times}  
\usepackage{helvet}  
\usepackage{courier}  
\usepackage[hyphens]{url}  
\usepackage{graphicx} 
\usepackage{natbib}  
\usepackage{caption} 
\usepackage{algorithm}
\usepackage{algorithmic}
\usepackage{amsmath}
\usepackage{bbding}
\usepackage{graphicx}
\usepackage{amsfonts}
\usepackage{subfigure}
\usepackage{newfloat}
\usepackage{listings}
\DeclareCaptionStyle{ruled}{labelfont=normalfont,labelsep=colon,strut=off} 
\floatstyle{ruled}
\newfloat{listing}{tb}{lst}{}
\floatname{listing}{Listing}
\title{SE Territory: Monaural Speech Enhancement Meets the Fixed Virtual Perceptual Space Mapping}
\author{
    Xinmeng Xu$^{1}$,
    Yuhong Yang$^{1}$,
    Weiping Tu$^{1,2*}$,
}
\affiliations{
    $^1$National Engineering Research Center For Multimedia Software, Wuhan University \\
    $^2$Hubei Luojia Laboratory\\

}

\usepackage{bibentry}

\begin{document}

\maketitle

\begin{abstract}
Monaural speech enhancement has achieved remarkable progress recently. However, its performance has been constrained by the limited spatial cues available at a single microphone. To overcome this limitation, we introduce a strategy to map monaural speech into a fixed simulation space for better differentiation between target speech and noise. Concretely, we propose SE-TerrNet, a novel monaural speech enhancement model featuring a virtual binaural speech mapping network via a two-stage multi-task learning framework. In the first stage,  monaural noisy input is projected into a virtual space using supervised speech mapping blocks, creating binaural representations. These blocks synthesize binaural noisy speech from monaural input via an ideal binaural room impulse response. The synthesized output assigns speech and noise sources to fixed directions within the perceptual space. In the second stage, the obtained binaural features from the first stage are aggregated. This aggregation aims to decrease pattern discrepancies between the mapped binaural and original monaural features, achieved by implementing an intermediate fusion module. Furthermore, this stage incorporates the utilization of cross-attention to capture the injected virtual spatial information to improve the extraction of the target speech. Empirical studies highlight the effectiveness of virtual spatial cues in enhancing monaural speech enhancement. As a result, the proposed SE-TerrNet significantly surpasses the recent monaural speech enhancement methods in terms of both speech quality and intelligibility.
\end{abstract}

\section{Introduction}
Speech enhancement (SE) stands as a crucial and formidable challenge in the realm of speech applications. Its core objective is to extract clean speech from noisy input signals \cite{ref1, ref2}. Occupying a pivotal role in various speech-related systems, SE finds applications spanning from speech recognition front-ends to hearing aids tailored for individuals with hearing impairments \cite{ref3, ref4,ref5}. SE methods can be classified into two distinct categories based on the number of sensors or microphones utilized: monaural (single-channel) and multi-channel (array-based) SE methods.

The monaural SE methods supply a simple but effective solution to the SE tasks by processing speech signals from a single microphone. In recent years, a plethora of monaural SE algorithms have emerged, falling into two primary categories: statistical-based algorithms and data-driven-based methods. The former encompasses techniques like spectral subtraction  \cite{ref6}, Wiener filtering \cite{ref7}, minimum mean-square error (MMSE) estimator \cite{ref8}, and the optimally modified log-spectral amplitude speech estimator \cite{ref9}. On the other hand, data-driven-based SE algorithms, primarily utilizing deep neural networks (DNNs), have gained prominence. DNN-based approaches treat SE as a supervised learning problem  \cite{ref10, ref11, ref12}, enabling them to learn discriminative patterns of speech and noise interference from training data. Despite their ability to capture speech features in both time and frequency domains, DNN-based methods still face limitations in performance, as extracting more discriminative patterns from single-channel recordings remains challenging~\cite{ref13, ref14}.

\begin{figure}
  \centering
  \includegraphics[width=0.78\linewidth]{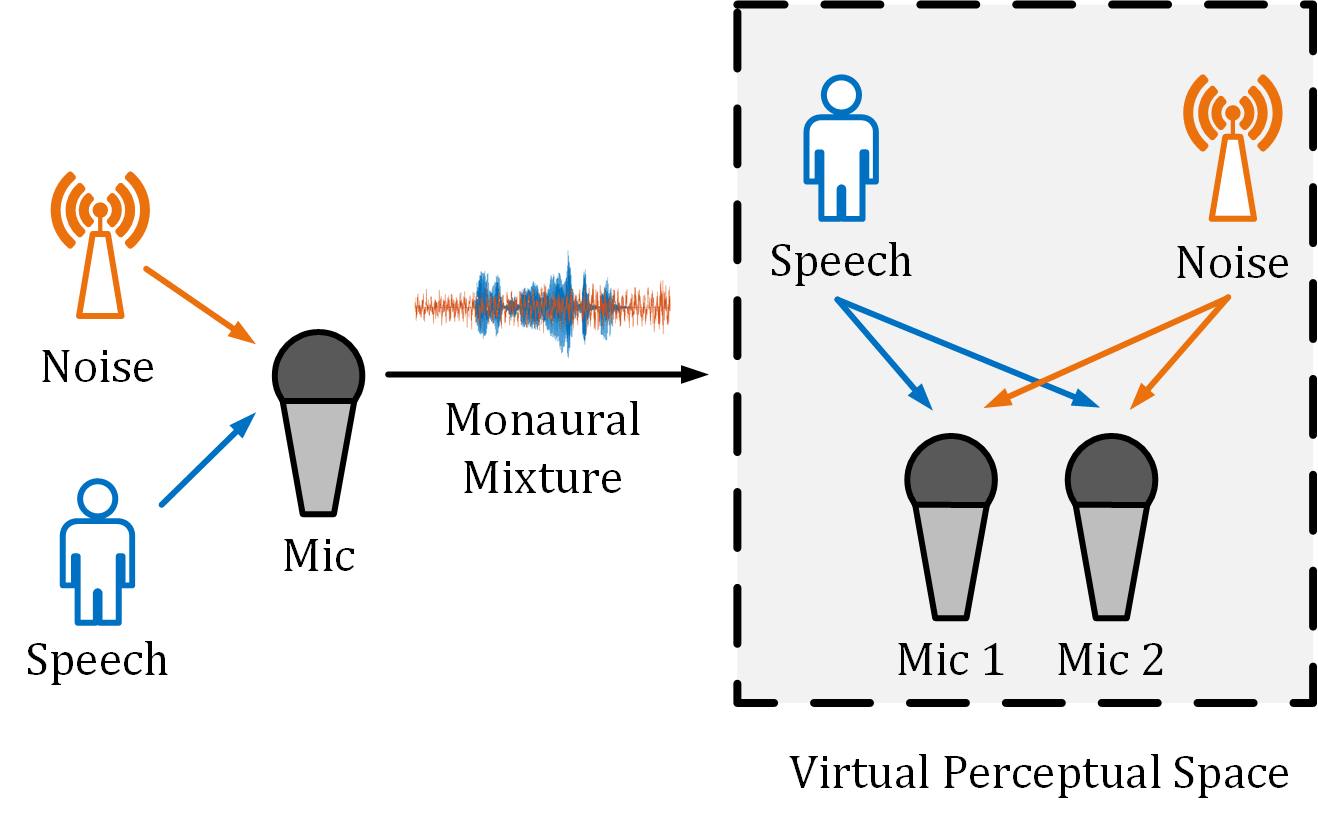}
  \caption{Demonstration of virtual perceptual space mapping, which aims to map monaural speech to a binaural speech, in which the desired speech and noise are set to be in an ideal and fixed direction that contains more discriminative spatial cues. We achieve virtual perceptual space mapping by utilizing an ideal binaural room impulse response (BRIR) originating from the corresponding monaural mixture.}
  \label{fig:1}
\end{figure}

The preceding discussion pertains exclusively to monaural SE tasks. However, exploiting spatial information can lead to further advancements in SE performance \cite{ref15,ref16,ref17}. With access to multiple microphones, both the spectral and spatial characteristics of speech and noise sources can be harnessed, leading to a procedure that integrates spatiotemporal information. Consequently, multi-channel SE frameworks exhibit reduced speech distortion compared to monaural SE methods, making them more suitable for real-world scenarios. Nonetheless, training a multi-channel SE model with robust generalization capabilities demands a significant number of noisy/clean speech pairs to effectively capture the variations in the spatial distribution features of audio signals \cite{ref18, ref19}. Meeting this rigorous requirement for training data may prove time-consuming and resource-intensive.

Through the above analysis, monaural and multi-channel have unique drawbacks but can complement each other in many aspects. For instance, monaural SE systems require only a single microphone to collect audio data, which is cost-effective and do not need a large amount of training data to cover the variations in spatial distribution features of noisy speech. But multi-channel SE systems can extract spatial information that is an effective cue to enhance the SE performance from their input signals. These motivate us to design a novel SE method, which takes the monaural speech signal as input while relishing the spatial information. In order to achieve monaural SE by using spatial information, two problems need to be solved, (1) How to produce the spatial information for a monaural speech without obeying the paradox that spatial information can be self-generated from the one-channel audio recording, and (2) how to fully extract the more discriminative patterns of speech and noises with the produced spatial information.

To address these challenges, we draw inspiration from spatial audio mapping techniques \cite{ref20, ref21, ref22} and introduce a strategy that involves mapping the monaural speech mixture into a predetermined virtual perceptual space to enhance speech enhancement performance. Illustrated in Figure~\ref{fig:1}, this virtual perceptual space is pre-established, where the intended speech and noise components are positioned in an ideal and fixed direction. To achieve this goal, we introduce a novel monaural SE model with a virtual binaural speech mapping network named SE-TerrNet. This is a two-stage network that is trained within a multi-task learning framework. Concretely, the first stage renders the monaural speech into the virtual perceptual space by sequentially employing several supervised speech mapping (SSM) blocks to progressively learn the binaural noisy speech synthesized from monaural input using an ideal type of binaural room impulse response. The second stage aggregates the binaural features from each SSM block in the first stage and extracts the spatial information by using the proposed intermediate fusion and cross-attention modules, respectively, to achieve the SE task. Note that the mapped speech in the virtual perceptual space only contains a single type of pre-defined spatial distribution feature, and the binaural room impulse response of the spatial distribution is a prior knowledge in our method, which avoids the paradox that monaural speech can self-generate spatial information.

This paper makes several notable contributions that are summarized as follows:

\begin{itemize}
    \item We introduce SE-TerrNet, a two-stage monaural SE algorithm that operates within a multi-task learning framework. This model incorporates spatial information through virtual perceptual space mapping.
    
    \item We propose novel modules: an intermediate fusion module and a cross-attention module. These modules serve to aggregate the binaural features generated by each SSM block in the initial stage and effectively capture spatial information respectively.
    
    \item To assess the efficacy of SE-TerrNet, we conduct comprehensive ablation studies using publicly available speech data. The results highlight that SE-TerrNet outperforms existing state-of-the-art methods. Notably, the integration of antiphasic binaural presentation space mapping enhances SE-TerrNet's superiority over alternative approaches.
\end{itemize}

\begin{figure}
\centerline{\includegraphics[width=0.8\linewidth]{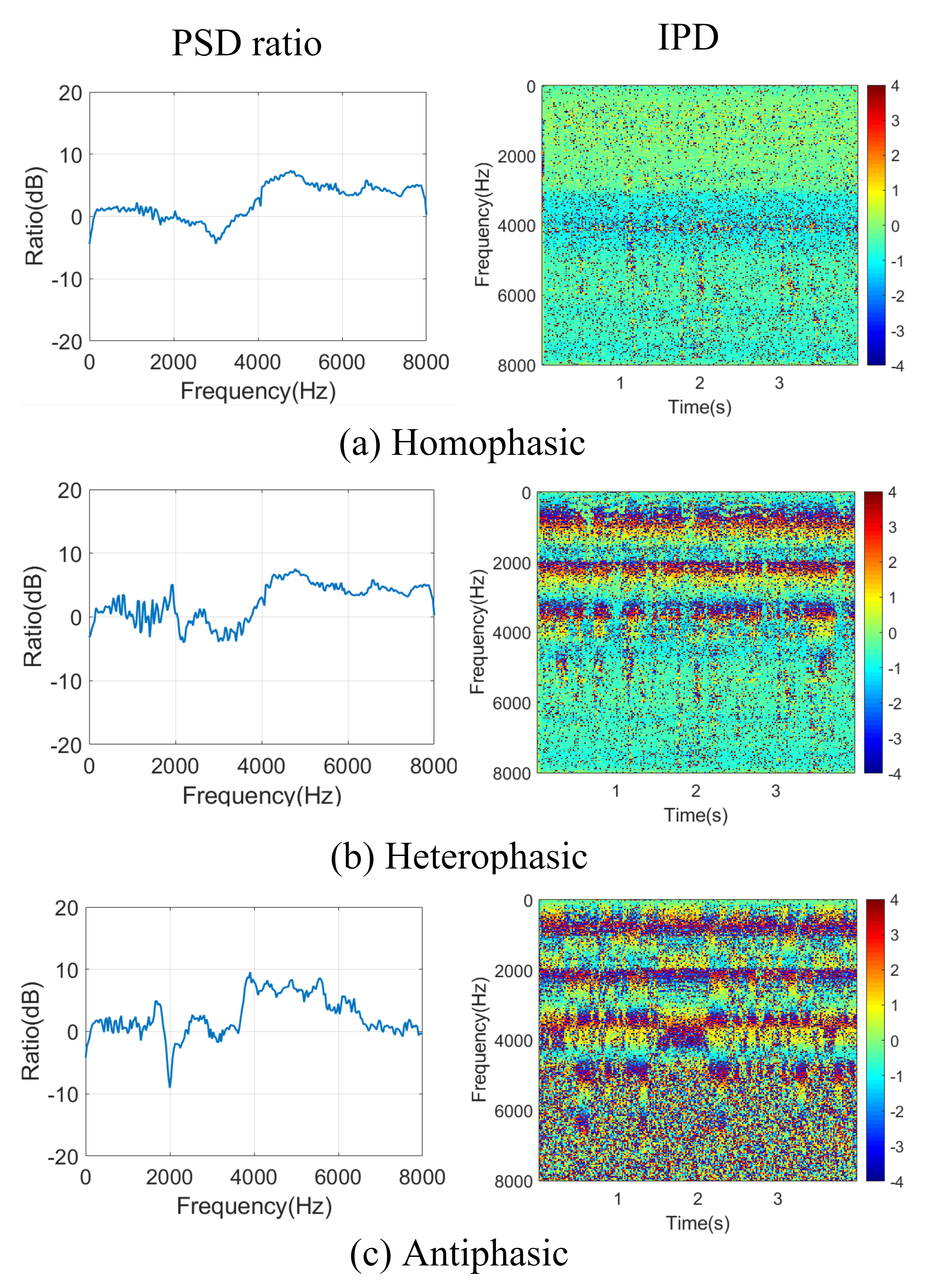}}
\caption{Illustrative instances of power spectral density (PSD) ratio and inter-channel phase difference (IPD) in distinct presentation scenarios: (a) Homophasic, (b) Heterophasic, and (c) Antiphasic.}
\label{fig:2}
\end{figure}

\begin{figure*}[t]
  \centering
  \includegraphics[width=0.9\linewidth]{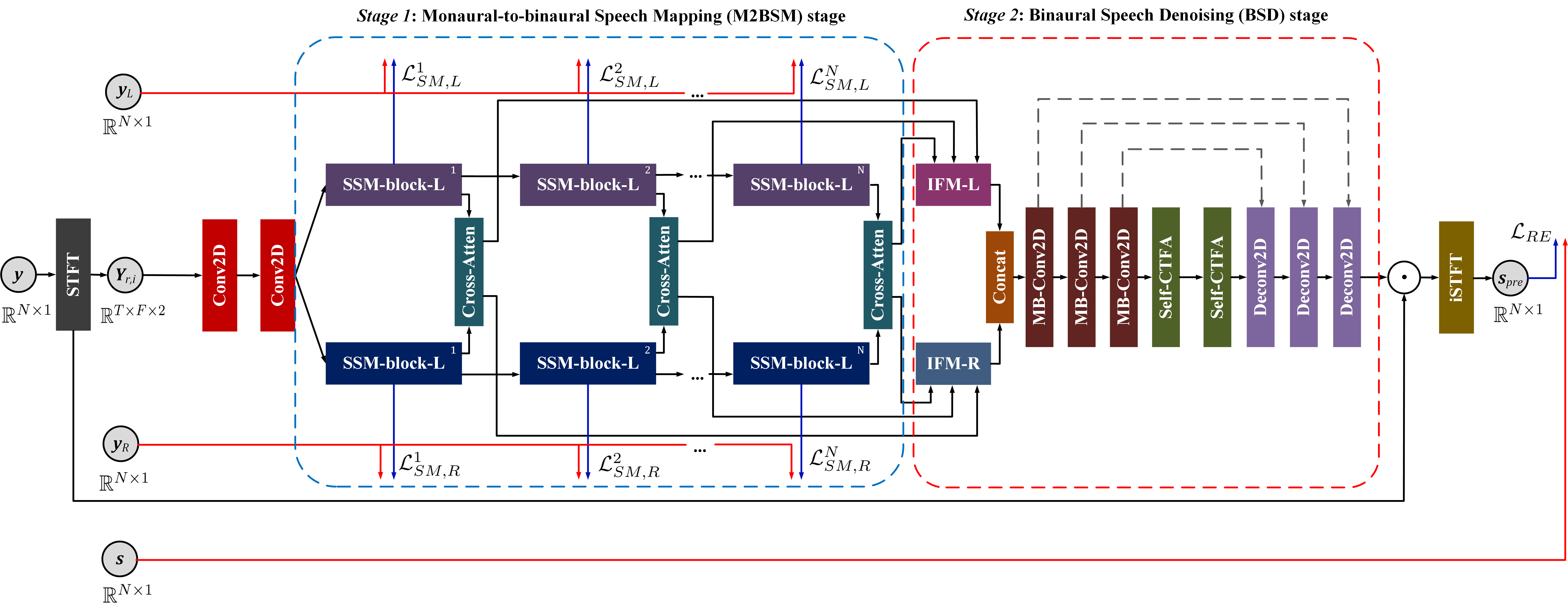}
  \caption{Illustration of the proposed SE-TerrNet framework for monaural speech enhancement.}
  \label{fig:3}
\end{figure*}

\section{Related Works}
\subsection{Monaural Speech Enhancement}
In the realm of monaural speech enhancement (SE), the central objective involves processing noisy speech from a solitary channel while directly aiming to eliminate noise components. Initial methodologies primarily centered on discerning speech and noise patterns within short time windows, a strategy termed local processing. Convolutional neural networks (CNNs) gained prominence in SE tasks due to their efficacy in handling local patterns like harmonics in speech spectrograms, yielding commendable performance in sequential modeling  \cite{ref24, ref25}. Nevertheless, speech is imbued with non-linguistic long-range dependencies, spanning attributes like gender, dialect, speaking style, and emotional state \cite{ref24, ref25}. To capture more informative speech features, diverse strategies have emerged, including recurrent neural networks (RNNs) \cite{ref27, ref28}, convolutional recurrent networks (CRNs) \cite{ref31, ref30}, and transformer neural networks \cite{ref32, ref33}. Although these frameworks excel in single-channel settings, they encounter challenges when transposed to real-world scenarios. Many of these methods introduce speech distortion to reduce noise, thereby compromising their generalizability. Additionally, they overlook the potential of harnessing spatial information, a key principle of the human auditory system. To summarize, while monaural SE methodologies have made notable strides, the demand persists for approaches that adeptly capture long-range dependencies, harness spatial information, and sustain robust performance in realistic environments.

\subsection{Binaural Presentation of Noisy Speech} \label{sec:2.3}
In the realm of psychoacoustic research \cite{ref40}, the binaural presentation of noisy speech has been categorized into three distinct scenarios: antiphasic, heterophasic, and homophasic. Notably, the antiphasic presentation, involving a $180^\circ$ phase difference between speech and noise, yields the highest level of speech intelligibility. Following this, the heterophasic presentation, encompassing speech and noise within a phase range of $0^\circ$ to $180^\circ$, is the next in line in terms of speech intelligibility. On the other hand, the homophasic presentation, where speech and noise are rendered in phase at $0^\circ$ (akin to monaural presentation), exhibits the lowest speech intelligibility. Furthermore, the power spectral density (PSD) ratio and inter-channel phase difference (IPD) of these three scenarios are visually depicted in Figure~\ref{fig:2}. It is notable that the binaural speech signals within the antiphasic presentation scenario exhibit the highest rate of change in PSD ratio and the most pronounced inter-channel phase difference. These observations correspond well with the principle that the antiphasic presentation leads to the highest speech intelligibility, surpassing both the heterophasic and homophasic (monaural) presentations \cite{ref20, ref40}. In our work, we effectively utilize the antiphasic binaural presentation to establish a virtual perceptual space, enhancing the clarity of spatial information within our research context.

\section{Proposed Scheme}
\subsection{Overview}
The basic idea behind SE-TerrNet is to render monaural noisy speech into an ideal and fixed virtal perceptual space,  \textit{i.e.}, antiphasic binaural presentation, for obtaining more discriminative spatial information to improve the performance of SE models. As shown in Figure~\ref{fig:3}, the proposed SE-TerrNet consists of 2 stages, (1) the monaural-to-binaural speech mapping (M2BSM) stage, including several SSM blocks, for progressively fixed virtual perceptual space mapping and (2) the binaural speech denoising (BSD) stage for noise reduction and target speech reconstruction.

Initially, the network receives an input in the form of $\textbf{Y}_{r, i}\in \mathbb{R}^{T\times F\times2}$, a complex-valued spectrogram obtained through short-time Fourier transform (STFT), where $T$ signifies the number of time steps, and $F$ represents the number of frequency bands. This $\textbf{Y}_{r,i}$ is then directed through consecutive 2D convolutional blocks. Each block comprises a 2D convolutional layer, followed by batch normalization and an exponential linear unit (ELU), yielding the intermediate feature $\textbf{Y}_c\in \mathbb{R}^{T\times F\times C_c}$ with $C_c$ denoting output convolution channels.

Subsequently, $\textbf{Y}_c$ is input to M2BSM stage. This stage is designed with two streams: $L$ for the left channel and stream $R$ for the right channel, for binaural speech mapping. These streams share the same network architecture but possess distinct parameters. Multiple SSM blocks are stacked within the M2BSM stage to gradually transform monaural input into the virtual perceptual space. A cross-attention block is introduced to enhance network performance by effectively utilizing spatial information from binaural data. The outputs of each SSM block, after being processed by cross-attention, are fed into intermediate fusion modules (IFMs) to aggregate features. These IFMs contribute to feature alignment, as the output of earlier SSM blocks aligns more with the input, while later SSM blocks contain more evident spatial information. The IFM outputs (IFM-L for the left channel and IFM-R for the right channel) are then concatenated and forwarded to the BSD stage. This stage follows an encoder-decoder framework for speech denoising and reconstruction.

Finally, the outcome $\textbf{S}_{out}$, achieved by multiplying the BSD stage output with $\textbf{Y}_{r, i}$, s fed into an inverse STFT (ISTFT) to convert the feature back into the enhanced time-domain signal,  $\textbf{s}_{pre} \in \mathbb{R}$. The SE-TerrNet, which is structured with two stages within a multi-task learning framework, is trained using a combination of two loss functions. To begin, the training objective for the M2BSM stage employs the signal-distortion index \cite{ref20}.For the $i^{th}$ left channel SSM block, this loss, denoted as follows:
\begin{equation}
    \mathcal{L}^i_{\textit{SM,L}}=10\log_{10}\Big\{ \frac{E\big\{[y_L(n)-\hat{y}^{(i)}_L(n)]^2 \big\}}{E[y^2_L(n)]} \Big\}.\label{eq:1}
\end{equation}
A corresponding signal-distortion index for the $i^{th}$ right channel SSM block, $l^i{sm-R}$, is defined in a similar manner as Equation~\ref{eq:1}. The simulation of the target binaural noisy speech involves the following expressions:
\begin{align}
    &y_L(n) =x(n)*h_{x,L}(n)+v(n)*h_{v,L}(n), \label{eq:2}\\
    &y_R(n) =x(n)*h_{x,R}(n)+v(n)*h_{v,R}(n). \label{eq:3}
\end{align}
Here, $x(n)$ and $v(n)$ represent scaled speech and noise, while $h_{x,L}(n)$, $h_{x,R}(n)$, $h_{v,L}(n)$, and $h_{v,R}(n)$ denote the acoustic impulse responses from the desired speech and noise rendering positions to the left and right ears. The M2BSM stage's training objective becomes a summation:
\begin{equation}
\mathcal{L}_{\textit{SM}}=\sum^{N}_{i=1}\Big(\mathcal{L}^i_{\textit{SM,L}}+\mathcal{L}^i_{\textit{SM,R}}\Big), \label{eq:4}
\end{equation}
The second loss focuses on training the BSD stage, minimizing the SI-SNR loss between the estimated clean speech $s_{pre}$ and the ground truth clean speech $s$, defined as $\mathcal{L}{RE}$. The final total loss is given by:
\begin{equation}
    \mathcal{L}_{\textit{total}} = \gamma \mathcal{L}_{\textit{SM}} + \mathcal{L}_{\textit{RE}},
\end{equation}
Due to the differing scales of these two losses, the parameter $\gamma$ is set empirically to 0.01.

\begin{figure}
\centerline{\includegraphics[width=0.9\linewidth]{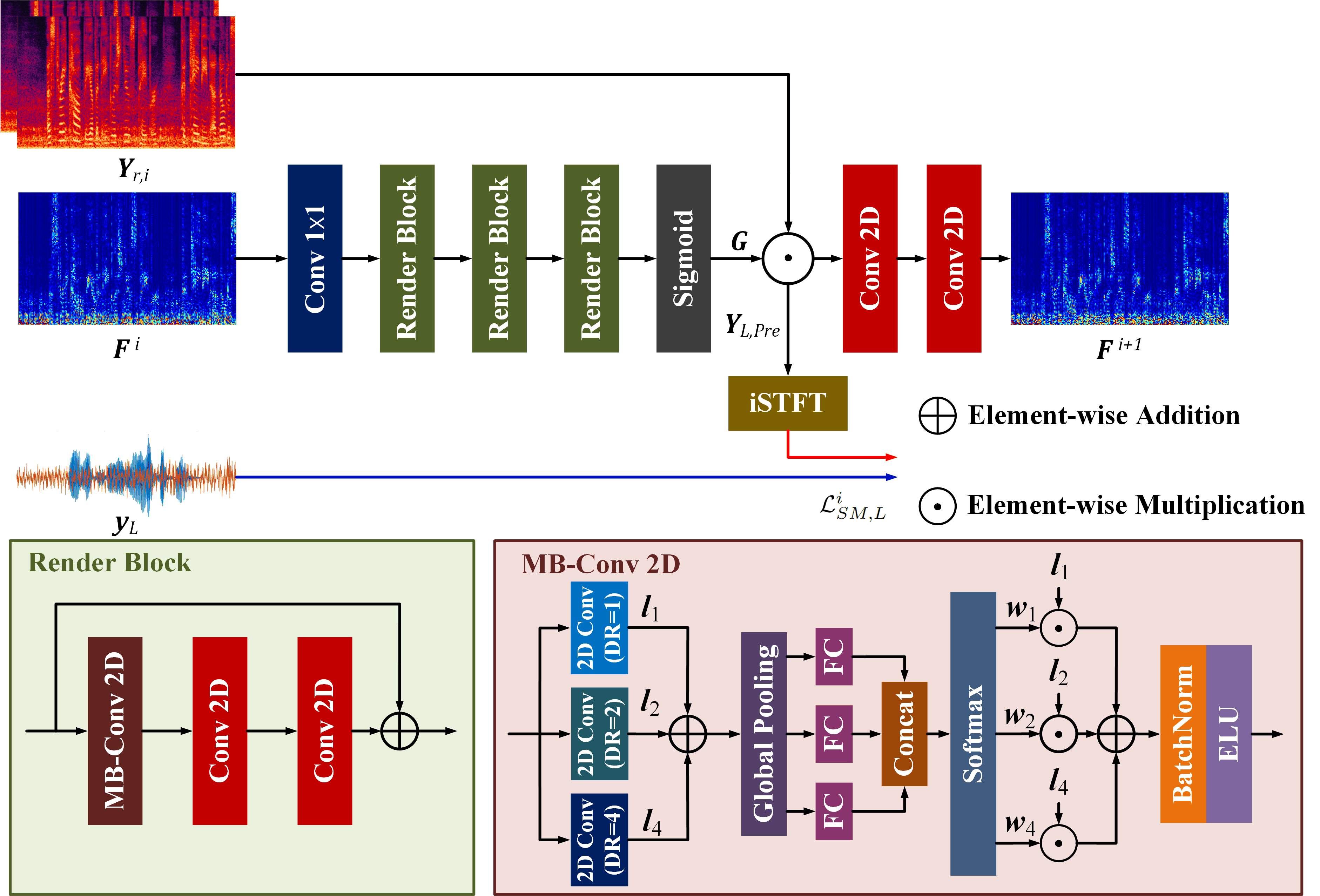}}
\caption{Block diagram of $i^{th}$ SSM block in the stream of the left channel.}
\label{fig:4}
\end{figure}

\subsection{Monaural to Binaural Speech Mapping Stage}
We design the M2BSM stage that contains several SSM block pairs for mapping the left and right channels of the binaural speech. The schematic diagram of $i$th left-channel SSM-block is shown in Figure~\ref{fig:4}. SSM takes the incoming feature $\textbf{F}^{i}$ from the previous layer as input and adopts a $1\times 1$ convolutional layer, several render blocks (3 is the default), and a sigmoid activation function to generate the rendering vector $\textbf{G}$. Next, $\textbf{G}$ is multiplied to the monaural noisy spectrogram to obtain the mapped (left-channel) spectrogram $\textbf{Y}_{L, Pre} \in \mathbb{R}^ {T\times F \times2}$. Then, an ISTFT is used to transform $Y_{L, pre}$ back to time-domain signal $\textbf{y}_{L,Pre}\in \mathbb{R}^{N\times 1}$. The supervision information is provided using simulated binaural signal, as Equation~\ref{eq:2} and~\ref{eq:3}, where the loss function is presented as Equation~\ref{eq:4}. Finally, the predicted virtual perceptual space mapped feature vector, $\textbf{F}^{i+1}$ is obtained from $\textbf{Y}_{L, Pre}$ processed by two 2D convolutional layers followed by a layer normalization and an ELU function.

\begin{figure}
\centerline{\includegraphics[width=0.85\linewidth]{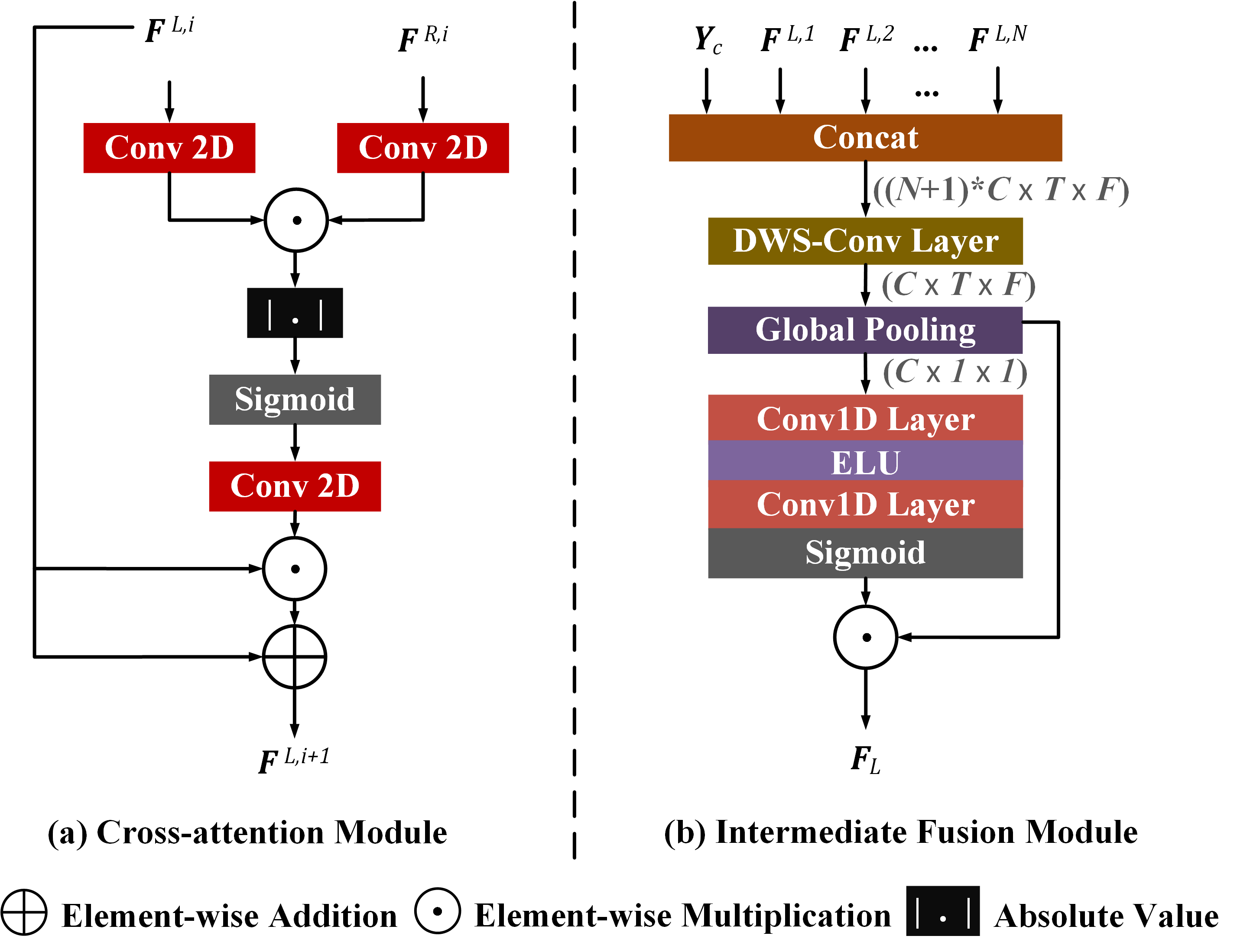}}
\caption{Block diagram of (a) $i^{th}$ cross-attention module in the stream of the left channel and (b) intermediate fusion module for left-channel features aggregation.}
\label{fig:5}
\end{figure}

The rendering block serves as the central component of the SSM block, comprising a multi-branch 2D convolutional (MB-Conv 2D) block and two additional convolution blocks. The MB-Conv 2D block, specifically designed for this approach, consists of three branches, each utilizing convolutional layers with varying dilation rates. This configuration enables the generation of feature maps with distinct receptive field sizes. Additionally, channel-wise attention is applied independently to the outputs of these three branches, and their results are combined. This approach allows for the extraction of features through a continuum of local to non-local operations, leading to increased flexibility. Moreover, it achieves enlargement of the receptive field without incurring significant computational overhead, thanks to the parallel structure of the three dilated convolutional layers employing the same kernel size \cite{ref41}. As a result, the rendering block optimizes the feature extraction process while effectively managing computational resources.

\begin{figure}
\centerline{\includegraphics[width=0.9\linewidth]{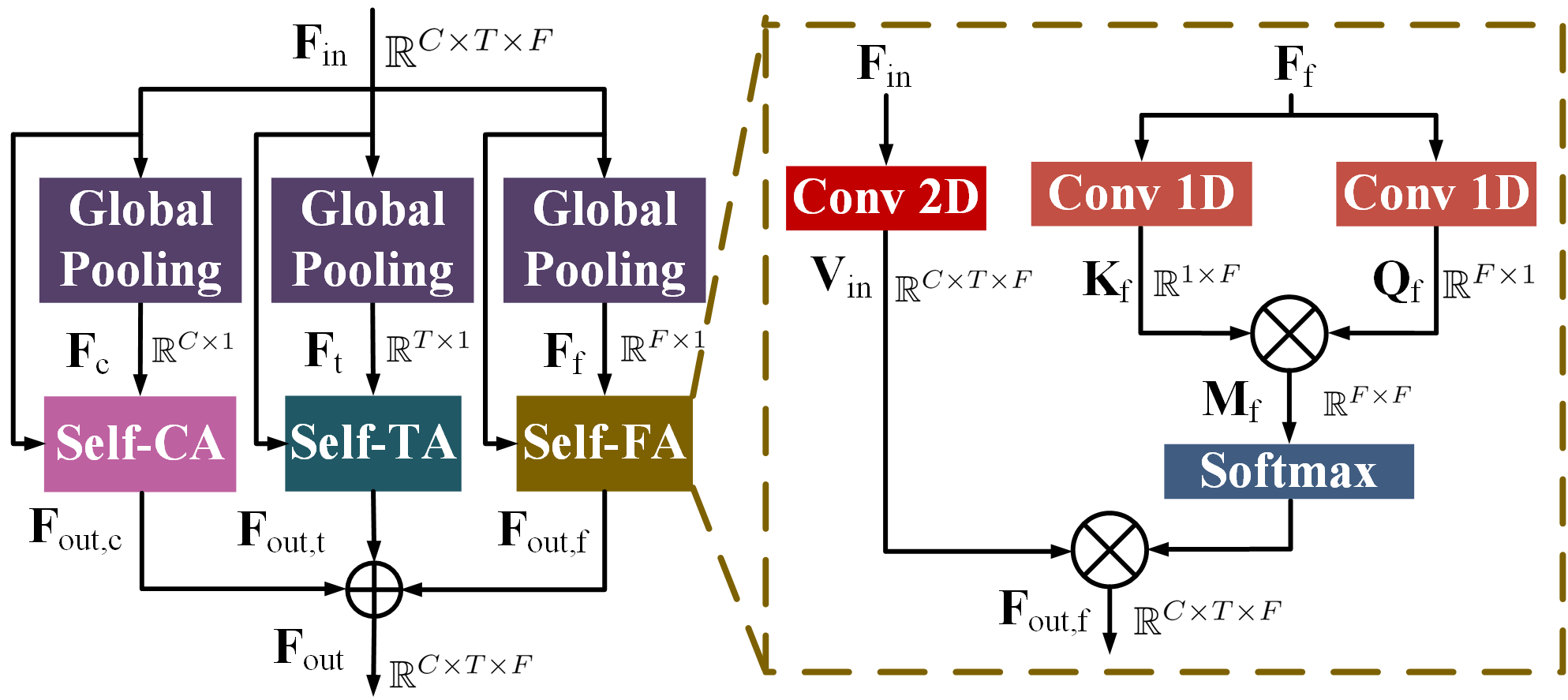}}
\caption{The architecture of Self Channel-Time-Frequency Attention (Self-CTFA) Module.}
\label{fig:6}
\end{figure}

\subsection{Binaural Speech Denoising (BSD) Stage}
The main objective of the BSD stage is to utilize spatial information for noise reduction and speech reconstruction. It takes the aggregated binaural features as input, generating the target RI-spectrograms, $\textbf{S}_{out}$. A cross-attention mechanism is designed to extract cross-channel features from each SSM block output pair. Furthermore, an intermediate fusion module (IFM) consolidates features from these cross-attention modules, forming the aggregated binaural features.

In Figure 5~\ref{fig:5} (a), cross-attention starts by passing binaural signal features through separate 2D convolutional layers (kernel size 1) with tanh activation, bounding inputs within [-1, 1] \cite{ref42}. The signals are then element-wise multiplied, emphasizing regions with gradual temporal variations and high power. The resulting mask is transformed back using 2D convolution (kernel size 1) and a sigmoid function. This mask is applied to the input feature, with a residual connection addressing gradient vanishing \cite{ref42}. This cross-attention mechanism facilitates extracting relevant information from binaural signals for further enhancement in the proposed method. In Figure~\ref{fig:5} (b), IFM first concatenates the features from all cross-attention with $\textbf{Y}_c$ and then feeds the concatenation into depth-wise separable (DWS) convolution layer for independent channel processing \cite{ref43}. Next, IFM adopts a channel attention module \cite{ref44} consisting of a global pooling operation and two 1-D convolutional layers adaptively select the channels of output feature of DWS-convolution to get the aggregated feature.

Aggregated binaural features from cross-attention and IFM are used in the BSD stage. The BSD stage concatenates features of the left and right channels, employing an encoder-decoder denoising network, in which three MB-convolution blocks are performed as the encoder, three deconvolutional blocks are performed as the decoder, and skip connections are inserted between the encoder and the decoder.  Two self-channel-time-frequency attention (Self-CTFA) modules enhance the bottleneck, as depicted in Figure~\ref{fig:6}. Each Self-CTFA module takes $\textbf{F}_{in} \in \mathbb{R}^{T\times F\times C}$ as input, generating energy distributions for channel, time, and frequency dimensions. Attention feature maps $\textbf{M}_c$, $\textbf{M}_t$, and $\textbf{M}_f$ are created through query-key interactions and softmax. Subsequently, feature vectors $\textbf{V}{in}$ are multiplied with attention maps. A feed-forward network with bidirectional GRU produces self-CTFA module outputs.

\section{Experiment}
\subsection{Datasets}
\textbf{Training Data:} The training dataset is formulated via a multi-step procedure integrating the WSJ0-SI 84 speech dataset \cite{ref45}, which encompasses 7138 utterances from 83 speakers (42 males and 41 females), the DNS-Challenge noise dataset \cite{ref46}, and selected binaural room impulse responses (BRIR) from MCIRs \cite{ref47}. These MCIRs utilize eight-microphone linear arrays featuring distinct inter-microphone spacings, while the BRIRs are recorded under three reverberation durations (0.16, 0.36, and 0.61 seconds). The process of generating training data entails three crucial phases:
\begin{itemize}
    \item \textbf{Trimming}: The length of chosen noise samples ($v(n)$) is adapted to align with the length of corresponding clean speech samples ($x(n)$), ensuring uniform durations for both noisy and clean speech.
    \item \textbf{Speech Level Scaling}: The speech intensity is adjusted within the range of -35 dB to -15 dB before noise introduction, as formulated:
    \begin{equation}
        \hat{x}(n)=\nu \cdot x(n),\label{eq7}
    \end{equation}
    where $\nu =10^{\epsilon/20}/\sigma_x$, $\epsilon$ is randomly chosen from the range $-35:1:-15$ dB, and $\sigma_x = \sqrt{E[x^2(n)]}$.
    \item \textbf{Noise Level Scaling}: Prior to addition to $\hat{x}(n)$, noise is scaled to control the signal-to-noise ratio (SNR), as defined by:
    \begin{equation}
        \hat{v}(n)=\vartheta \cdot v(n), \label{eq8}
    \end{equation}
    where $\vartheta=10^{\frac{-\texttt{SNR}}{10}\sigma^2_{\hat{x}}/\sigma^2_v}$, $\sigma^2_{\hat{x}}=E[\hat{x}^2(n)]$, $\sigma^2_v=E[v^2(n)]$, and SNR is randomly selected from the range $-15:1:-15$ dB.
\end{itemize}

\begin{figure*}[t]
\centerline{\includegraphics[width=0.85\linewidth]{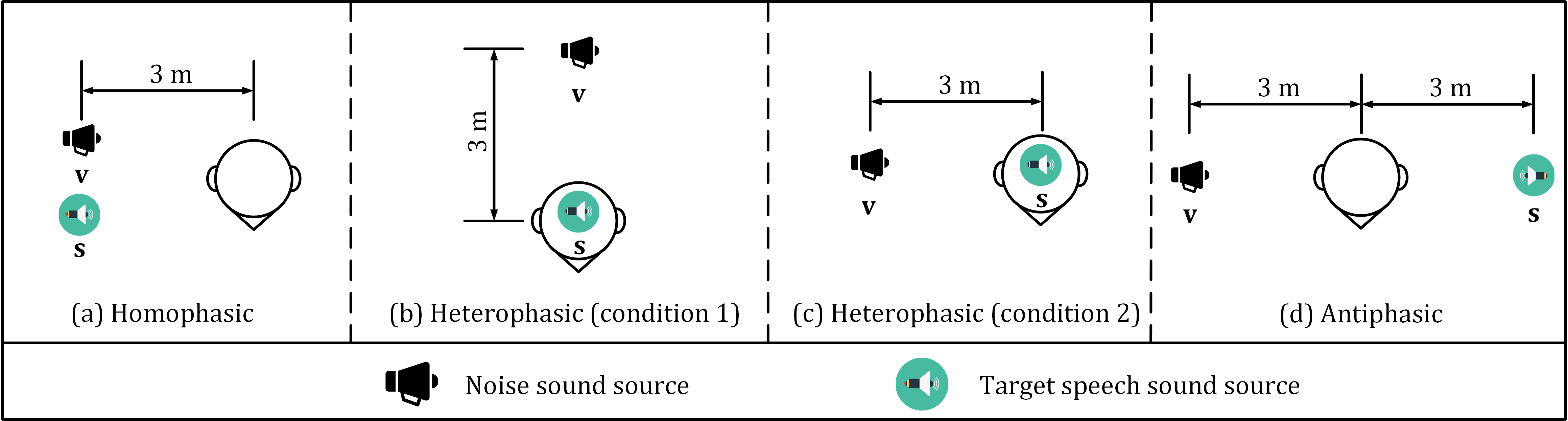}}
\caption{Illustration of 4 different direction perceptions of binaural presentation.}
\label{fig:7}
\end{figure*}

Post these three phases, the noisy monaural speech is generated as:
\begin{equation}
    y(n) = \hat{x}(n) + \hat{v}(n).
\end{equation}
Additionally, the signals of the left and right channels for noisy binaural speech are generated using Equations~\ref{eq:2} and ~\ref{eq:3}.

\textbf{Testing Data:} To generate the test dataset, we solely require the speech and noise datasets. Our experimentation involves the evaluation of SE-TerrNet on two distinct datasets:
\begin{itemize}
    \item \textbf{Voice Bank + DEMAND} \cite{ref48}: This dataset consists of 840 utterances sourced from 84 speakers in the Voice Bank dataset \cite{ref49}. The dataset is augmented with unfamiliar varieties of \textit{non-stationary} noises extracted from DEMAND \cite{ref50}.

    \item \textbf{AVSpeech + AudioSet}: In this dataset, we merge 800 utterances from the AVSpeech dataset \cite{ref51} with randomly selected noise excerpts from AudioSet \cite{ref52}. The linear combination is defined as $\text{Mix}_i = \text{Speech}_j + 0.3 \times \text{Noise}_k$, where $\text{Speech}_j$ and $\text{Noise}_k$ denote 4-second segments randomly chosen from the speech and noise datasets, respectively.
\end{itemize}

\subsection{Implementation Details}
Within our experimental framework, we gauge the effectiveness of our approach through the assessment of two key objective metrics: short-term objective intelligibility (STOI) \cite{ref53} and perceptual evaluation of speech quality (PESQ) \cite{ref54}—both of which play a pivotal role in our ablation study. Moreover, for a comprehensive model evaluation, we incorporate CSIG \cite{ref56}, CBAK \cite{ref56}, and COVL \cite{ref56} as additional performance metrics. Enhanced scores across these metrics are indicative of improved audio quality. To implement our method, we harness the PyTorch framework. Our preprocessing involves resampling all audio samples to a consistent 16kHz frequency. For the computation of the short-time Fourier transform (STFT), a Hann window is utilized, featuring a length of 25ms, a hop length of 10ms, and an FFT size of 512. This configuration results in an input audio feature representation of dimensions $251 \times 257 \times 2$.

\subsection{Ablation Study}
\renewcommand{\arraystretch}{0.95}
\begin{table}[]
\centering
\begin{tabular}{c|cc|c}
\hline
\# of SSM block pairs & STOI (\%) & PESQ & Params. \\ \hline
1                    & 94.27    & 3.09 & 4.20 M        \\
2                    & 94.85    & 3.16 & 5.26 M       \\
3                    & 95.71    & 3.25 & 6.31 M       \\
4                    & 95.94    & 3.27 & 7.37 M       \\
5                    & 96.08    & 3.27 & 8.43 M       \\ \hline
\end{tabular}
\caption{Ablation study on the number of SSM block pairs.}
\label{tab:1}
\end{table}

\noindent \textbf{Number of SSM block pairs.} \quad The SSM block pairs in the M2BSM stage are designed for progressively mapping the monaural speech into the pre-defined virtual perceptual space. To trade off the STOI and PESQ scores of SE-TerrNet between different SSM block pairs in the M2BSM stage, we train the SE-TerrNet with varying numbers of SSM block pairs in the M2BSM stage. Table\ref{fig:1} shows the results, indicating that larger models perform better. Using 3 SSM block pairs yields significant gains of 0.18 PESQ and 1.44$\%$ STOI compared to 1 learning block. However, 4 SSM blocks show only marginal improvement with more parameters. We choose 3 SSM block pairs as the final model, striking a good balance between computation costs and accuracy for comparison with other methods.

\renewcommand{\arraystretch}{0.95}
\begin{table}[t]
\centering
\begin{tabular}{l|cc}
\hline
Binaural Presentation & PESQ & STOI(\%) \\ \hline
Homophasic            & 2.98 & 92.52    \\ 
Heterophasic (condition 1)       & 3.11 & 93.14    \\ 
Heterophasic (condition 2)       & 3.12 & 93.06    \\ 
Antiphasic            & 3.25 & 95.71    \\ \hline
\end{tabular}
\caption{Ablation study of binaural presentation}
\label{tab:2}
\end{table}

\noindent \textbf{Impact of different binaural presentation.} \quad In theory, distinct binaural presentation conditions yield varying degrees of discriminative spatial information. The antiphasic presentation, where speech and noise are rendered in opposite directions, provides the most discriminative spatial cues. Conversely, the heterophasic presentation situates speech centrally in the listener's head and noise on the sides. In contrast, the homophasic presentation delivers the least discriminative spatial information since both speech and noise are perceived in the same region. To accommodate these various presentation conditions, we trained four networks accordingly, as depicted in Figure~\ref{fig:7}. The outcomes presented in Table~\ref{tab:2} validate that our approach aligns with the fundamental principles of binaural presentation for noisy speech. This also underscores the efficacy of the M2BSE-Net in mapping monaural speech into the predefined virtual perceptual space while effectively utilizing spatial information to enhance overall performance. By incorporating diverse binaural presentation conditions, we optimize the utilization of discriminative spatial cues, resulting in improved speech enhancement outcomes.

\renewcommand{\arraystretch}{0.95}
\begin{table}[t]
\centering
\resizebox{0.47\textwidth}{!}{
\begin{tabular}{l|cccccc}
\hline
Case Index      & 0 & 1     & 2     & 3     & 4     & 5     \\ \hline
MB-Conv         & $\checkmark$         & $\times$   & $\times$   & $\checkmark$   & $\checkmark$   & $\checkmark$   \\
Conv (DR=1)   & $\times$         & $\checkmark$   & $\times$   & $\times$   & $\times$   & $\times$   \\
Conv (DR=4)   & $\times$         & $\times$   & $\checkmark$   & $\times$   & $\times$   & $\times$   \\
Self-CTFA       & $\checkmark$         & $\checkmark$   & $\checkmark$   & $\times$   & $\checkmark$   & $\checkmark$   \\
SA  & $\times$         & $\times$   & $\times$   & $\checkmark$   & $\times$   & $\times$   \\
CA & $\checkmark$         & $\checkmark$   & $\checkmark$   & $\checkmark$   & $\times$   & $\checkmark$   \\
IFM             & $\checkmark$         & $\checkmark$   & $\checkmark$   & $\checkmark$   & $\checkmark$   & $\times$   \\ \hline
STOI (\%)        & 95.71       & 94.84 & 94.74 & 96.04 & 94.27 & 94.33 \\
PESQ            & 3.25        & 3.17  & 3.16  & 3.27  & 3.18  & 3.13  \\
Param. (M)      & 6.31        & 6.38  & 6.31  & 6.94  & 5.64  & 5.85  \\ \hline
\end{tabular}}
\caption{Ablation study on different components of BSD stage. ``DR'' indicates dilation rate, ``SA'' indicates self-attention, and ``CA'' indicates cross-attention.}
\label{tab:3}
\end{table}

\noindent \textbf{Impact of different components in the BSD stage.} \quad We present the validation results in Table~\ref{tab:3} to investigate the effectiveness of different components within the BSD stage, focusing on PESQ, STOI, and the number of parameters. Notably, \textbf{Case 0} represents the basic BSD stage with default settings. In this analysis, we contrast the proposed BSD stage with various modules: (1) \textbf{Case 1}: replacing the MB-Convolution with a convolution block featuring a dilation rate of 1, (2) \textbf{Case 2}: replacing the MB-Convolution with a convolution block having a dilation rate of 4, (3) \textbf{Case 3}: replacing the self-CTFA with self-attention (SA) \cite{ref57}, (4) \textbf{Case 4}: removing the cross-attention module, and (5) \textbf{Case 5}: discarding IFMs and concatenating features from the last SSM block pairs processed by cross-attention into the denoising network.

\renewcommand{\arraystretch}{1}
\begin{table}[t]
\resizebox{0.47\textwidth}{!}{
\begin{tabular}{lccccc}
\hline
\textbf{Method} & \textbf{PESQ} & \textbf{CSIG} & \textbf{CBAK} & \textbf{COVL} & \textbf{Param.} \\ \hline
Noise           & 1.97          & 3.35          & 2.44          & 2.63          & -               \\
SEGAN           & 2.16          & 3.48          & 2.94          & 2.80          & -               \\
DEMUCS          & 3.07          & 4.31          & 3.40          & 3.63          & 60.80 M         \\
DCCRN           & 2.68          & 3.88          & 3.18          & 3.27          & 3.67 M          \\
TFT-Net         & 2.75          & 3.93          & 3.44          & 3.34          & 5.81 M          \\
PHASEN          & 2.99          & 4.21          & 3.55          & 3.62          & 6.83 M          \\
SN-Net          & 3.12          & 4.39          & 3.60          & 3.77          & 8.14 M               \\
FAF-Net         & 3.19          & 4.13          & 3.38          & 3.66          & -               \\ \hline
SE-TerrNet      & \textbf{3.25}         & \textbf{4.58}          & \textbf{3.71}          & \textbf{3.98}          & 6.31 M          \\ \hline
\end{tabular}}
\caption{Model Comparison on VoiceBank + DEMAND.}
\label{tab:4}
\end{table}

Comparing \textbf{Case 0}, \textbf{Case 1}, and \textbf{Case 2}, we note that the convolution block with a dilation rate of 1 performs comparably to the MB-convolution block, albeit with higher computational complexity. Conversely, the convolution block with a dilation rate of 4 maintains a lower computational complexity, albeit at the cost of performance. Comparing \textbf{Case 0} and \textbf{Case 3}, we observe that the self-CTFA module significantly improves system performance while exhibiting notably lower computational complexity than SA. Lastly, the comparisons between \textbf{Case 0} and both \textbf{Case 4} and \textbf{Case 5} underscore the indispensability and effectiveness of the proposed cross-attention module and IFM.

\renewcommand{\arraystretch}{0.95}
\begin{table}[t]
\centering
\begin{tabular}{lccc}
\hline
\textbf{Method} & \textbf{SSNR (dB)} & \textbf{PESQ} & \textbf{STOI(\%)} \\ \hline
Conv-TasNet     & 13.26              & 2.92          & 83.14            \\
DCCRN           & 13.94              & 2.89          & 85.83           \\
PHASEN          & 15.89              & 3.04          & 87.94            \\
U-Former        & 15.34              & 3.06          & 88.87            \\
CASE-Net        & 16.47              & 3.11          & 89.33            \\ \hline
SE-TerrNet      & \textbf{17.24}     & \textbf{3.19} & \textbf{89.69}   \\ \hline
\end{tabular}
\caption{Model comparison on AVSpeech + AudioSet.}
\label{tab:5}
\end{table}

\subsection{Model Comparison}
\textbf{VoiceBank + DEMAND.} In our assessment using the compact yet widely recognized dataset, we performed a direct comparative analysis of the proposed SE-TerrNet against seven cutting-edge techniques. Among these, three adopt U-shaped network architectures, namely SEGAN \cite{ref58}, DEMUCS \cite{ref59}, and DCCRN \cite{ref31}, while the remaining four employ cylindrical network architectures, encompassing TFT-Net \cite{ref60}, PHASEN \cite{ref10}, SN-Net \cite{ref12}, and FAF-Net \cite{ref61}. The outcomes, detailed in Table~\ref{tab:4}, unequivocally demonstrate that SE-TerrNet outperforms all other methods across all evaluation metrics, all the while maintaining a favorable computational complexity. Importantly, the results for the selected baseline models are directly extracted from their respective research papers, ensuring a fair and dependable comparison. The exceptional performance of SE-TerrNet on this dataset serves as a robust validation of its stature as a top-tier speech enhancement approach.

\textbf{AVSpeech + AudioSet.}  We conducted a comparison of our method with five other recent approaches on this large dataset. The compared methods include Conv-TasNet \cite{ref62}, DCCRN \cite{ref31}, PHASEN \cite{ref10}, U-former \cite{ref32}, and CASE-Net \cite{ref35}. Conv-TasNet is a time-domain method, DCCRN and U-Former are two U-shaped networks, and CASE-Net is a monaural SE model by leveraging speech local and non-local features. For a fair comparison, we train these baseline models on WSJ0-SI 84 + DNS Challenge datasets and evaluate them on AVSpeech + AudioSet datasets. The results in Table~\ref{tab:5} demonstrate that the proposed SE-TerrNet outperforms all these five methods. The outstanding performance exhibited on a large dataset underscores the generalizability of our method to diverse speakers and a wide range of noisy environments. 

\section{Conclusion}
In this paper, we propose a novel SE-TerrNet for monaural SE. The proposed SE-TerrNet aims to overcome the challenge that spatial information is unavailable in monaural speech signals. To address this issue, we map the monaural speech signals into the pre-defined virtual perceptual space. In particular, we design SE-TerrNet following a 2-stage framework. The first stage maps the monaural speech into a virtual perceptual space by sequentially employing several SSM block pairs that allow for progressively achieving binaural signal mapping. The second stage performs speech denoising and reconstruction based on binaural speech representation inputs with meticulously designed cross-attention and IFM. It is important to highlight that binaural signal mapping is achieved by utilizing a binaural room impulse response (BRIR) within an antiphasic binaural presentation originating from the corresponding monaural mixture. Our experimental results demonstrate the superiority of the proposed method over prior arts.

\bibliography{aaai24}

\end{document}